# Schottky barrier and attenuation length for hot hole injection in non-epitaxial Au on *p*-type GaAs

I. Sitnitsky, J. J. Garramone, J. R. Abel, and V. P. LaBella[a)]

College of Nanoscale Science and Engineering, University at Albany, SUNY, Albany, New York 12203

P. Xu, S. D. Barber, M. L. Ackerman, J. K. Schoelz, and P. M. Thibado[b)]

Department of Physics, University of Arkansas, Fayetteville, Arkansas 72701

Ballistic electron emission microscopy (BEEM) was performed to obtain nanoscale current versus bias characteristics of non-epitaxial Au on *p*-type GaAs in order to accurately measure the local Schottky barrier height. Hole injection BEEM data was averaged from thousands of spectra for various metal film thicknesses and then used to determine the attenuation length of the energetic charge carriers as a function of tip bias. We report the marked increase in attenuation length at biases near the Schottky barrier, providing evidence for the existence of coherent BEEM currents in Schottky diodes. These results provide additional evidence against the randomization of a charge carrier's momentum at the metal-semiconductor interface.

Electronic mail: [a)]vlabella@albany.edu; [b)]thibado@uark.edu

## I. INTRODUCTION

Metal-semiconductor (MS) Schottky barrier diodes are vital to modern electronics. The transport of energetic charge carriers at the MS interface in nanoscale Schottky diodes is most suitably studied with ballistic electron emission microscopy (BEEM),[1-11] a three terminal scanning tunneling microscopy (STM) and spectroscopy technique introduced in the late 1980s. In a BEEM experiment, biased charge carriers are injected into the grounded metal base of the Schottky diode, and charges that are sufficiently energetic to overcome the Schottky barrier are registered as the BEEM current. Accurate measurement of this barrier height $\phi_B$ is of fundamental importance for engineering devices which utilize Schottky diodes, but the task is made difficult by a sensitive dependence on spatial inhomogeneities at the MS interface. Fortunately, the BEEM tip allows the subsurface interface to be probed with atomic-scale spatial resolution, while the narrow energy distribution of the injected charge carriers results in a high energetic resolution of about 0.02 eV.

In the ideal theoretical description of BEEM currents,[2] it is assumed that the charge carrier's momentum parallel to the MS interface, $k_\parallel$, and its energy are both conserved. In addition, the energy and momentum distributions at the MS interface are presumed to be identical to those at injection. Attempts to confirm or disprove this model, however, have been inconclusive. For example, if the momentum distribution is narrow, a delay of about 1 eV is expected in the current onset for Si(111) with respect to the onset for Si(001)[12] because at energies near the conduction band minima, the electronic structure of Si(111) does not have states available with zero parallel momentum.[13] Yet, in experiments with non-epitaxial MS systems, no delayed onset[12,13] or only a shallow onset[7] has been observed. Thus one could alternatively propose that $k_\parallel$ is not conserved at the interface owing to, for instance, elastic momentum randomization events.[12,14] In fact, it is possible in this way to reproduce from theory the aforementioned similar onset biases, but such a model is in conflict with the high spatial resolution obtained by BEEM.[3-9] Furthermore, evidence against momentum randomization has been provided by epitaxial MS systems[15,16] as well as recent BEEM experiments which demonstrate a sharp increase in the attenuation length of the charge carriers in Ag/Si(001) for energies near $\phi_B$ but not in Ag/Si(111).[17]

In this paper we present BEEM measurements of $\phi_B$ for non-epitaxial Au on *p*-type GaAs. Au/GaAs has been extensively studied as a prototype for non-epitaxial metal interfaces on direct band gap semiconductors. Our technique uses an automated scanning system that allows for a large number of local measurements to be taken across the sample surface and averaged. In addition, the attenuation length $\lambda$ of the energetic charge carriers is measured by varying the Au thickness, which provides insight into charge transport through the diode. A sharp increase in the observed $\lambda$ as tip biases approach $\phi_B$ indicates that $k_\parallel$ is likely conserved in charge transport across the MS interface, in accordance with the recent results from similar experiments using Ag on *n*-type Si.

## II. EXPERIMENTS

The GaAs sample was prepared in a liquid-nitrogen cooled, ultra-high vacuum (~3 × $10^{-10}$ mbar) molecular beam epitaxy (MBE) growth chamber (Riber 32), which includes a dual Ga effusion cell, a two-zone As valved-cracker cell, and a reflection high-energy electron diffraction (RHEED) system operating at 15 keV. A commercially available, "epiready," 2 in. diameter, $p^+$ (Zn-doped $10^{19}$ cm$^{-3}$) GaAs(001) ± 0.1° wafer was indium mounted on a 2 in. diameter, standard MBE molybdenum block and loaded into the MBE chamber without chemical cleaning. To remove the surface oxide layer, the substrate was heated to 580 °C while exposing the surface to $10^{-5}$ mbar As$_4$. Before growth, the Ga and As$_4$ beam equivalent pressures were set and measured to give an As$_4$/Ga flux ratio of 10. The substrate temperature was then maintained at 580 °C, and GaAs was grown for 4 hours. RHEED oscillations determined the growth rate of GaAs to be ~1.3 $\mu$m/hr, giving a final sample thickness of ~5 $\mu$m. Although the film was undoped, MBE-grown GaAs films are slightly *p*-type due to As antisite defects (~$10^{15}$ cm$^{-3}$).[18] Following growth, the chamber was cooled, and the GaAs sample was removed from the chamber and transferred to our metals deposition facility to create Schottky diodes.

The native oxide formed during transfer was removed from the film using a standard chemical treatment immediately prior to loading it into the high vacuum deposition chamber.[9,19] Au films 19, 21, and 24 nm thick were deposited through a 1 mm$^2$ shadow mask using a Varian electron-beam deposition system with a base pressure of $10^{-7}$ mbar. More thicknesses were attempted; however, thinner Au did not result in a continuous film, and thicker gold resulted in BEEM currents that were too small. After deposition, the sample was mounted onto a custom designed sample holder for BEEM measurements that allowed for simultaneous grounding of the metal film using a BeCu contact and connection of the GaAs to the *ex situ* pico-ammeter to measure the BEEM current. The ohmic contact to the GaAs was fabricated by cold pressing indium into the backside of the GaAs wafer.

A modified low temperature, ultra-high vacuum STM (Omicron) with a pressure in the $10^{-11}$ mbar range was utilized for all BEEM measurements.[20] The samples were inserted into the chamber and loaded onto the STM stage that was cooled to 80 K for all BEEM measurements. Au STM tips that were mechanically cut at a steep angle were utilized for all measurements. For each sample, BEEM spectra were acquired from thousands of unique locations throughout a 3 $\mu$m × 3 $\mu$m area of the metal surface, using positive (hot hole injection) tip biasing conditions and a constant tunneling current setpoint of 5.0 nA. The BEEM spectra were then averaged to reduce the effects of surface roughness, which was determined to be about 1 nm from the STM images. The thickness of the metal films was determined *ex situ* by Rutherford backscattering spectrometry measurements.

For BEEM measurements with hole injection, which is much less common than electron injection, the tip is positively biased while the Au is held at ground, as shown schematically in Fig. 1(a). As in STM, the tip can move in three dimensions, and the current through the tip is monitored as it injects holes into the Au. Simultaneously, a pico-ammeter connected to the GaAs measures the BEEM current which has crossed the MS interface. An energy diagram illustrating the hot hole injection process is shown schematically in Fig. 1(b). The tip and the Au are separated by a vacuum barrier, which is typically modeled as trapezoid. A positive bias is applied to the STM tip, lowering its Fermi energy an amount $V_{tip}$ below that of the grounded Au film. Thus holes in the STM tip are shot into the metal. The semiconductor is *p*-type, so its valence band starts out just below the Fermi level. At the MS interface, however, the valence band maximum bends below the potential of the metal, forming a Schottky barrier and creating a barrier to hot hole injection.

## III. RESULTS

The BEEM spectra for the three different Au film thicknesses are shown in Fig. 2. While their shapes are similar, the overall magnitude of the BEEM current clearly decreases as the thickness of the Au layer increases. Each spectrum begins at a tip bias of -2.0 eV, where the hole transmission rate is ~1% for the 19 nm film, ~0.3% for 22 nm, and ~0.013% for 24 nm. As the tip bias goes to zero, the transmission also monotonically decreases to zero. The inset in the lower right-hand corner of the graph magnifies the 24 nm curve to better reveal its form. The Schottky hole barrier height, $\phi_B$, is marked on this spectrum at -0.5 eV.

For four characteristic tip biases close to $\phi_B$, the absolute value of the percent transmission ($I_{BEEM}/I_{tip} \times 100\%$) is plotted as a function of the Au film thickness in Fig. 3. A log scale is used on the *y*-axis, and a linear fit is applied to each data set. Each line is labeled with its corresponding tip bias, as well as the magnitude of its inverse slope, known as the attenuation length $\lambda$. Two noteworthy features of this plot are that the percent transmission decreases with increasing Au thickness, and that the slope increases ($\lambda$ decreases) with increasing bias.

The attenuation length is plotted for all tip biases in Fig. 4. It starts at approximately 1.1 nm for a tip bias of -2.0 eV and slowly grows to 1.28 nm at -0.86 eV. The attenuation then dips to 1.27 nm at -0.76 eV before rapidly increasing, reaching 2.8 nm at the Schottky hole barrier height, -0.5 eV. The energy diagram of a hole being injected directly at the top of the Schottky barrier is displayed as an inset in the center of Fig. 4. As before, the Fermi energy of the Au is shown as a dashed line across the diagram, and the valence band maximum in the GaAs bends below the Fermi level at the MS interface. However, the tip bias has been adjusted to equal $\phi_B$ so that the injected hole depicted in this illustration is at the height of the valence band as it contacts the MS interface.

**IV. DISCUSSION**

The energy level schematic shown in Fig. 1(b) is one of the keys to understanding this experiment. The difference between the Fermi level and the top of the GaAs valence band at the interface is defined as $\phi_B$. A positive bias placed on the tip lowers the energy of the injected holes below the Fermi level of the grounded Au film, but only as this bias is increased past $\phi_B$ will some holes will be sufficiently energetic to cross the barrier into the semiconductor, registering as the BEEM current $I_{BEEM}$. Thus the Schottky barrier height could be estimated from the spectra in Fig. 2 by simply noting the tip bias at which there is an onset of BEEM current. More rigorously, it may be ascertained by fitting the curves to $I_{BEEM}/I_{tip} \propto |V_{tip} - \phi_B|^{2.5}$, where $I_{tip}$ is the tunneling current.[2,21,22] Using this method, $|\phi_B|$ was determined to be $0.50 \pm 0.02$ eV for Au/$p$-GaAs at 80 K. We are confident in this measurement due to the statistical averaging employed in our procedure. In addition, our results for Au on $n$-type GaAs (not presented) give $|\phi_B| = 0.96 \pm 0.02$ eV, so the sum is 1.46 eV, consistent with the band gap of GaAs at 80 K.

The points represented in Fig. 3 can be extracted from Fig. 2 by drawing vertical lines through the spectra at the four chosen tip biases. Unsurprisingly, the percent transmission is found to decrease with increasing Au thickness[8] due to an increased probability of scattering in the metal film. Each data set at a given tip bias is individually fit to

$$I_{BEEM}/I_{tip} \propto e^{-\frac{d}{\lambda(E,T)}}, \tag{1}$$

where $I_{BEEM}$ is the current at the collector, $I_{tip}$ is the tip or base current, $d$ is the film thickness, and $\lambda(E,T)$ is the hot hole attenuation length, which depends on the energy $E$ and temperature $T$.[9] Thus one may demonstrate that the slope of the resulting line is inversely proportional to $\lambda$, and the decreasing slopes as the bias approaches $\phi_B$ must indicate an increase in $\lambda$. The sharp nature of this increase can be seen with $\lambda$ plotted versus tip bias, as in Fig. 4. At higher negative biases, $\lambda$ is relatively constant and equal to approximately 1.1 nm. The small local dip that occurs at $V_{tip} \approx -0.75$ eV can be attributed to an increase in $d$-band scattering from $d$-electrons at that energy below the Fermi level in Au, as observed in previous BEEM spectra from Cu/Si(001).[9,23-25] Between -0.7 and -0.5 eV, however, $\lambda$ more than doubles, increasing to a maximum of 2.8 nm.

The uptick can be explained by considering the momenta of charge carriers as they traverse the metal. For tip biases close to the barrier height, only forward moving ballistic charges will have enough energy in the proper direction to overcome the Schottky barrier and contribute to the BEEM current.[17] Therefore, the measured attenuation length at these biases will be particularly sensitive to the carriers with $k_\parallel \approx 0$. In contrast, at higher negative biases a broader range of carriers will now have enough forward momentum to contribute to the BEEM current, effectively providing a weighted average measurement for the observed $\lambda$. Very similar results were obtained in identical experiments performed on Ag/$n$-Si,[17] except the BEEM currents and attenuation lengths presented here for Au/GaAs are much smaller due to the interface band structure. It should be noted that, like Si(001) and unlike Si(111), GaAs has states available for charges with $k_\parallel \approx 0$, leading to our observation of a sharp increase in $\lambda$ near $\phi_B$. These results provide support for the existence of focused and coherent ballistic current in BEEM experiments performed on Schottky diodes and contradict the model of momentum randomization at the MS interface.

**V. CONCLUSION**

In conclusion, BEEM experiments were performed on custom-made nanoscale MS diodes. The Schottky barrier height for hot hole injection into *p*-type GaAs with Au contacts was determined to be -0.5 ± 0.02 eV at 80 K. A prominent spike in the attenuation length of the charge carriers for tip biases near $\phi_B$ provides evidence that the BEEM current in that regime consists of a small number of focused forward-moving charges that remain coherent as they cross the metal film.

**ACKNOWLEDGEMENTS**

The Albany group is thankful for the support of the Interconnect Focus Center, one of five research centers funded under the Focus Center Research Program, a DARPA and Semiconductor Research Corporation program. The Arkansas team gratefully acknowledges support from the National Science Foundation (NSF) under grant number DMR-0855358 and the Office of Naval Research (ONR) under grant number N00014-10-1-0181.

**Captions**

FIG. 1. (a) Schematic for the BEEM experimental setup for hot hole injection into Au on *p*-type GaAs. A special BeCu point contact is made to the Au film, while an indium contact was made to the GaAs using a cold press technique. (b) The corresponding energy diagram, illustrating the tip bias, hole injection, and band bending in the GaAs. A hole with sufficient energy to pass under the Schottky barrier before rising up to the Fermi level is shown.

FIG. 2. BEEM spectra showing percent transmission of the BEEM current divided by the tip current is shown versus tip bias for three different Au film thicknesses. Spectra were obtained with the STM feedback on to keep the tip current constant at 5 nA. Inset magnifies the curve associated with the 24 nm Au layer. The label just below the curve points to the tip bias at which the onset of the BEEM current is found.

FIG. 3. Percent transmission as a function of Au thickness plotted on a semi-log scale for four tip biases close to the Schottky barrier height. Solid lines are fits to Eq. (1), where the corresponding attenuation lengths and tip biases are indicated.

FIG. 4. Hot hole attenuation length versus tip bias. A dramatic increase is seen as the bias approaches $\phi_B$. The inset shows a potential diagram for a hole injected into the diode directly at the Schottky barrier height.

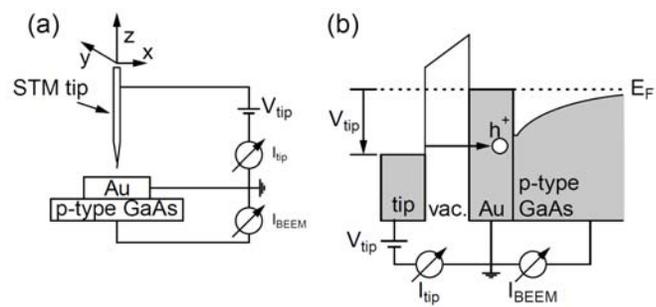

FIG. 1.

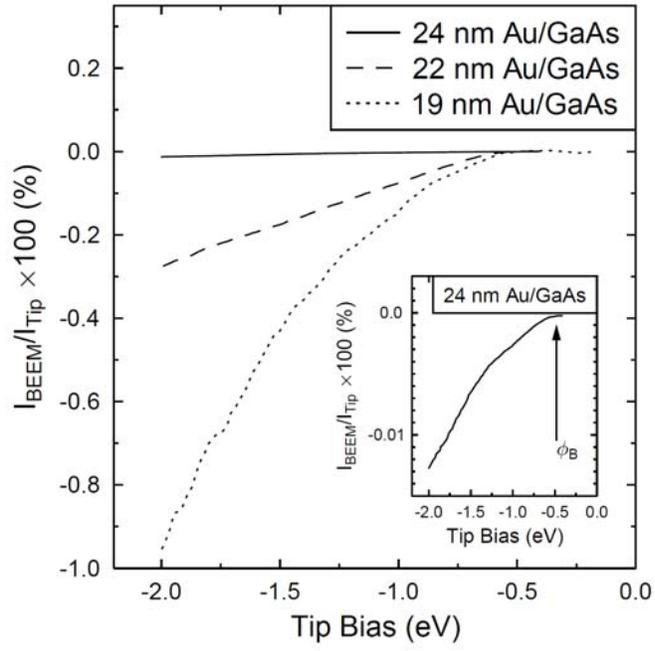

FIG. 2.

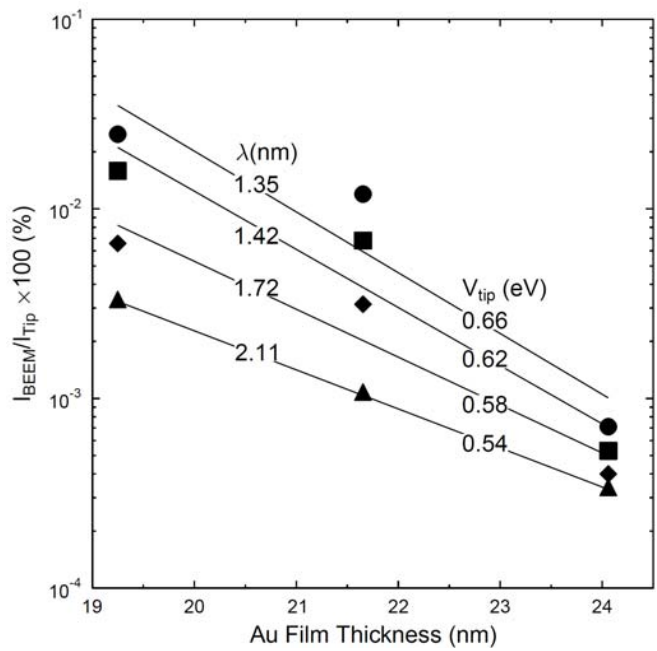

FIG. 3.

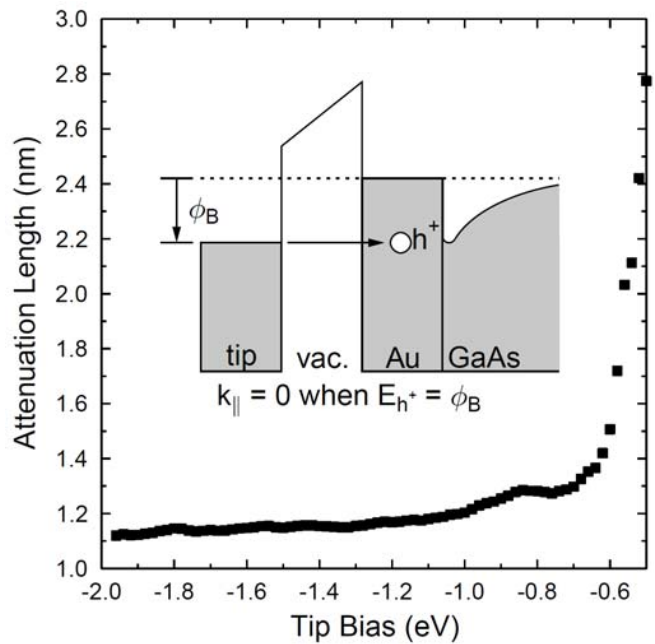

FIG. 4.